\documentstyle[sprocl]{article}

\input{psfig.sty}

\newcommand{\beba}{\begin{equation}\begin{array}{lcl}}
\newcommand{\eaee}{\end{array}\end{equation}}
\newcommand{\bea}{\begin{eqnarray}}
\newcommand{\eea}{\end{eqnarray}}
\newcommand{\ba}{\begin{array}}

\def\be{\begin{equation}}
\def\ee{\end{equation}}
\def\bea{\begin{eqnarray}}
\def\eea{\end{eqnarray}}

\newcommand{\sla}{\kern -5.4pt /}
\newcommand{\slaar}{\kern -7. pt / \kern 3.pt}
\newcommand{\Dir}{\kern -6.4pt\Big{/}}
\newcommand{\Dirin}{\kern -10.4pt\Big{/}\kern 4.4pt}
\newcommand{\DDir}{\kern -7.6pt\Big{/}}
\newcommand{\DGir}{\kern -6.0pt\Big{/}}

\newcommand{\beanon}{\begin{eqnarray*}}
\newcommand{\eeanon}{\end{eqnarray*}}
\newcommand{\ea}{\end{array}}
\newcommand{\ed}{\end{description}}
\newcommand{\bi}{\begin{itemize}}
\newcommand{\ei}{\end{itemize}}
\newcommand{\ben}{\begin{enumerate}}
\newcommand{\een}{\end{enumerate}}
\newcommand{\bc}{\begin{center}}
\newcommand{\ec}{\end{center}}

\def\epem{$e^+ e^-$\ }

\def\app #1 #2 #3 {{\it  Acta Phys.Polon.} {#1} (#2) #3\ }
\def\ap #1 #2 #3 {{\it Ann. Phys. }{ #1} (#2) #3\ }
\def\intj #1 #2 #3{{\it Int. J. Mod. Phys.} {#1} (#2) #3\ }
\def\hpa #1 #2 #3{{\it Helv. Phys. Acta. }{ #1} #2) #3\ }
\def\mpl #1 #2 #3 {{\it Mod.~Phys.~Lett.} {#1} (#2) #3\ }
\def\np #1 #2 #3 {{\it Nucl.~Phys.} {#1} (#2) #3\ }
\def\pl #1 #2 #3 {{\it Phys.~Lett.} {#1} (#2) #3\ }
\def\pr #1 #2 #3 {{\it Phys.~Rev.} {#1} (#2) #3\ }
\def\prep #1 #2 #3 {{\it Phys.~Rep.} {#1} (#2) #3\ }
\def\prl #1 #2 #3 {{\it Phys.~Rev.~Lett.} {#1} (#2) #3\ }
\def\rmp #1 #2 #3 {{\it Rev. Mod. Phys.} {#1} (#2) #3\ }
\def\zp #1 #2 #3 {{\it Z.~Phys.} {#1} (#2) #3\ }
\def\epj #1 #2 #3 {{\it Eur.~Phys.~J.} {#1} (#2) #3\ }
\def\cpc #1 #2 #3 {{\it Comp. Phys. Commun.} {#1} (#2) #3\ }
\def\xx #1 #2 #3 {{#1}, (#2) #3\ }

\begin{document}

\begin{flushright}

CTP-TAMU-36/99\\
November 1999\\
\end{flushright}

\vspace{2. truecm}

\title{
Four and Six Fermion Event Generators for $e^+e^-$ Collider
Physics{\footnote{To appear in the Proceedings of the International
Workshop on Linear Colliders, Sitges, Spain, April 28- May 5, 1999}}} 

\author{Elena Accomando}

\address{Phys. Dept.  Texas A \& M, College Station \\TX
77843, USA}

\maketitle\abstracts{
\vspace{1.5 truecm}
{\bf{Abstract:}} The status of four and six fermion event generators for
Standard
Model processes at present and future $e^+e^-$ colliders is briefly
reviewed.}

\vfill\eject

\title{
Four and Six Fermion Event Generators for $e^+e^-$ Collider Physics}

\author{Elena Accomando}

\address{Phys. Dept.  Texas A \& M, College Station \\TX
77843, USA}

\maketitle\abstracts{ 
The status of four and six fermion event generators for Standard
Model processes at present and future $e^+e^-$ colliders is briefly
reviewed.}

\section{Introduction}
The ongoing run at LEP2 has led in the last recent years to detailed
analyses related to both SM and beyond the SM physics \cite{yr}. The
theoretical results achieved for LEP2 physics have been implemented in
dedicated four fermion codes \cite{eg}. $M_W$ measurements, studies
of triple gauge couplings and Higgs searches in the low mass range require 
the analysis of four fermion final states, making therefore unavoidable
the calculation of four fermion processes at LEP2 and future $e^+e^-$
colliders. Since the advent of LEP2, several codes with different features
have been implemented and carefully cross-checked. The available four
fermion
tree level programs, most part now interfaced to QCD Generators 
(PYTHIA/JETSET and/or HERWIG), reviewed by Sj\"ostrand, to relate parton 
level predictions with experiments, are listed in Tab.[1].

The work done for four fermion physics at LEP2 can be extended to the
NLC. This machine however would not only improve the sensitivity to
trilinear gauge couplings and gauge boson properties, but it would be
also ideal for precision studies of quartic gauge couplings, top
properties and Higgs searches in the intermediate mass range \cite{rr}.
All these topics involve processes with six fermions in the final state,
but despite of this only a few six fermion codes have been at present
implemented. To our knowledge, only GRACE\cite{kuri}, SIXPHACT\cite{noi} 
and WWGENPV/ALPHA\cite{pv} can simulate six fermion processes. 

In this talk we summarize some features of the above
mentioned codes, even if this brief description can do no justice to the
effort that has been done by the various groups.

\section{Four fermion codes}
On the side of four fermion codes, many improvements have been recently
introduced into the existing programs (e.g. refer to contributions by
Jadach
\cite{koralw} and Perret-Gallix) and two new codes, presented by
Denner\cite{denner} and Peskin, have been implemented. 

Due to the increasing statistics at LEP2, the interplay between theory and
experiments is becoming more stringent. The attention is therefore
focusing on the role of the radiative effects and on the estimate of the
theoretical uncertainty on the predictions.
\begin{table}
\vspace{0.1cm}
\begin{center}
\unitlength 1cm
\begin{tabular}{|c|c|c|c |}
\hline
Code & processes & $m_f$ & AC \\
\hline\hline 
ALPHA & all & Y & N   \\
\hline
CompHEP & all & Y &  Y \\
\hline
ERATO & CC11/CC20 & N & Y  \\
\hline
EXCALIBUR & all & Y/N & Y  \\
\hline
GENTLE/4fan & CC11/NC & Y/N & Y  \\
\hline
grc4f & all & Y & Y  \\
\hline
KORALW & all & Y & Y  \\
\hline
LEPWW & CC3/NC2 & N & Y \\
\hline
PANDORA & CC3/NC2 &  & N \\
\hline
PYTHIA6/JETSET & New added & Y/N & N \\
\hline
WPHACT & all & Y & Y \\
\hline
WTO & all & N & N \\
\hline
WWGENPV/HIGGSPV & all & Y/N & N \\
\hline
WWHP & CC3 & & Y \\
\hline
\end{tabular}
\caption{Available four fermions programs. The included processes are
written using the notation of Ref.[1]. Y(N) indicates the code 
includes (does not) fermion masses ($m_f$) and anomalous couplings (AC).
Y/N is for approximate treatment of $m_f$.} 
\end{center}
\end{table}
An example is the treatment of the photons in the initial and final
state, a particularly delicate issue. Neglecting the $p_T$ of the photon,
as in
the usual LL approximation of the ISR, can have sizeable effects, for
example on the detection efficiency and on the differential distributions
used for TGC studies and also for New Physics searches since the detection
of new particles often rely upon missing $p_T$. Different prescriptions
have been adopted. Some codes include models for the generation of IS
multiphotons with finite $P_T$ (KORALW \cite{koralw}) and LL FSR,
while others
make use of QED parton showers with $p_T$ (EXCALIBUR, grc4f, LEPWW and
PYTHIA). A more accurate approach, not affected by possible gauge
invariance problems, could consist in merging the explicit emission of
a single hard photon with collinear bremsstrahlung and radiative
corrections. As a further step
toward the full calculation of the radiative effects,     
complete matrix elements for the process $e^+e^-\rightarrow 4f+\gamma$
have been implemented by ALPHA and recently by Denner et al.\cite{denner}.

Matching theory and experiments means also being able to produce reliable
results in a finite time. This question has focused the attention on the
integration methods. Since a large number of diagrams with different
topologies contribute to a given final state, the multichannel techniques
(as in EXCALIBUR, KORALW and WWGENPV/ALPHA) seem to be the most promising.

At the higher NLC energies, not only the complexity of the processes will
increase, but new difficulties could appear.
In Fig.[1], it is shown the result of one of the latest tuned comparisons
among four codes, concerning the gauge invariance issue for the single W
production at small angle, reviewed by Boos \cite{boos}. 
\begin{figure}[h]
\vspace{0.1cm}
\begin{center}
\unitlength 1cm
\begin{picture}(5.2,6.8)
\put(-1.8,-2.){\includegraphics{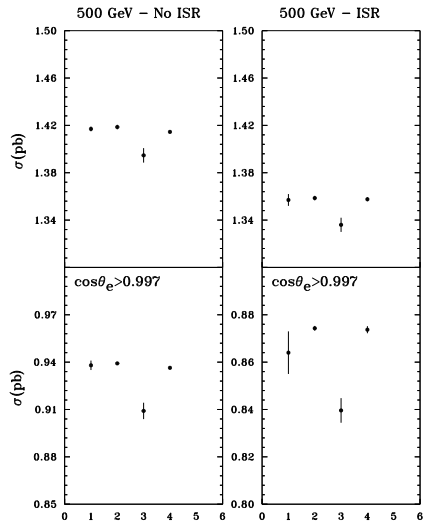}}
\end{picture}
\end{center}
\vspace{0.1cm}
\caption[]{
Comparison for single W production at small angle at 500 GeV with and
without ISR. On the X-axis: 1=CompHEP, 2=grc4f, 3=KORALW, 4=WPHACT. Cuts:
$M(u\bar d )\ge 5$ GeV and $E_(u,\bar d)\ge 3$ GeV. }
\label{h8}
\end{figure}
Events with one
or two electrons in the forward direction (coming from $e\nu W$, $eeZ$ or
$2\gamma$ processes) will be relevant at the NLC as signal, to disantagle
the $WW\gamma$ coupling, and as background to New Physics searches. These
processes, which contain t-channel photon diagrams with mass singularity
in the very forward region, require massive matrix elements and phase
space to cure these apparent divergencies. Moreover, they need a proper
implementation of the boson width in the propagators, in order to preserve
the large cancellations among the t-channel diagrams dictated by the gauge
invariance.
The comparison in Fig.[1] involves gauge preserving prescriptions, namely
the overall prescription (CompHEP and WPHACT) and  the $L_{\mu\nu}$
transform method (grc4f and KORALW), and shows a rather
satisfactory agreement. Recently, also a new scheme based on
complex gauge boson masses and obeying the Ward identities 
has been introduced, as presented by Denner \cite{denner}.

\section{Six fermion codes}

Six fermion final states  receive contributions from a great amount
of different Feynman diagrams. Some of them correspond to Higgs production
and decay. Others may come from $t\bar t$ or $WWZ$, or from partial
resonant processes (as for instance $\nu_e \bar \nu_e WW$) as well as non
resonant ones. All these diagrams, which in general are of the order of
several hundreds, correspond to the same final state and interfere among
themselves. Different resonant contributions might be considered as signal
or background depending on the process under study and their interplay can
be consistently analyzed only by complete calculations. Moreover, all
these channels might constitute a main source of background to New Physics
searches.  

Recently three groups\cite{kuri}\cite{noi}\cite{pv} have started
computing 
full tree level cross sections for six fermion final states at the NLC. 
These  computations have been applied to study the phenomenology of 
$t\bar t$\cite{kuri}\cite{noi}\cite{pv}, $WWZ$\cite{noi}\cite{pv} and 
higgs\cite{noi}\cite{pv} production and have already shown the importance
of 
 finite width effects and irreducible background. In these analyses,
the beamstrahlung effects have been simulated using the parametrization
implemented in CIRCE \cite{circe}. Recently, also a new method
(PYBMS), based on the Pisin-Chen formalism, has been proposed
\cite{pisin}.

\section{Conclusions}
Much effort has already been devoted in order to have reliable tools for
four fermion physics, but much work has still to be done on the side of
electroweak and QCD radiative effects. Whereas the four fermion area is
well covered, so far a few codes have been implemented for six fermion
processes. Hopefully, the joint work of the various groups will lead to a
better understanding of what is still missing in order to fully exploit
the great and unique capability of the $e^+e^-$ colliders.

\section*{Acknowledgments}
It is a real pleasure to thank the organizers for the nice and stimulating
atmosphere of the workshop. We wish to aknowledge also A. Ballestrero, E.
Boos, S. Jadach and D. Perret-Gallix for supplying us with
Fig.1 and for useful discussions.

\vfill\eject

\end{document}